\def\lte{\lower 0.5ex\hbox{${}\buildrel<\over\sim{}$}} 
\def\gte{\lower 0.5ex\hbox{${}\buildrel>\over\sim{}$}} 
\documentstyle[12pt,aasms]{article}
\input psfig.sty
\psfull

\begin{document}
\begin{center}
\title{Discovery of Pulsed X-ray Emission from the SMC Transient RX J0117.6-7330}
\vskip 5pt
D.J. Macomb$^{1,2}$,M.H. Finger$^{2,3}$,B. A. Harmon$^{3}$,R.C. Lamb$^{4}$,T.A. Prince$^{4}$
\end{center}

\affil{$^{1}$Lab. for High-Energy Astrophysics, NASA/GSFC, Greenbelt, MD 20771}
\affil{$^{2}$Astrophysics Program, University Space Research Association}
\affil{$^{3}$Space Sciences Laboratory, ES 84, NASA/Marshall Space Flight Center, Huntsville, AL 35812}
\affil{$^{4}$Space Radiation Laboratory, California Institute of Technology ,Pasadena, CA 91125}

\begin{abstract}
We report on the detection of pulsed, broad-band, X-ray emission from
the transient source RX J0117.6-7330.  The pulse period of 22 seconds
is detected by the ROSAT/PSPC instrument in a 1992 Sep 30 - Oct 2
observation and by the CGRO/BATSE instrument during the same epoch.
Hard X-ray pulsations are detectable by BATSE for approximately 100
days surrounding the ROSAT observation (1992 Aug 28 - Dec 8).  The
total directly measured X-ray luminosity during the ROSAT observation
is $1.0 \times 10^{38}$ (d/60 kpc)$^{2}$ ergs s$^{-1}$.  The pulse 
frequency increases rapidly during the outburst, with a peak spin-up
rate of $1.2 \times 10^{-10}$ Hz s$^{-1}$ and a total frequency change
of 1.8\%. 
The pulsed percentage is 11.3\% from 0.1-2.5 keV,
increasing to at least 78\% in the 20-70 keV band.  These results
establish RX J0117.6-7330 as a transient Be binary system.

\end{abstract}

\keywords{X-Rays:general --- Accretion: --- accretion disks: --- stars:individual(RX J0117.6-7330)}

\section{Introduction}
 
The PSPC instrument onboard the ROSAT spacecraft made an 8.98 ksec
observation of the Small Magellanic Cloud on 1992 Sep 30 - Oct 2
leading to the discovery (Clark, Remillard \& Woo 1996) of RX
J0117.6-7330, a bright X-ray source within 5 arc minutes of
SMC X-1.  The source was not detected in an observation of the
SMC a year earlier and was found 246 days later to be dimmer by 
more than 2 orders of
magnitude.  Further analysis showed that the X-ray luminosity of $2.3
\times 10^{37}$ (D/60 kpc)$^{2}$ ergs s$^{-1}$ (0.2-2.5 keV) was
derived assuming a position in the SMC (Clark, Remillard \& Woo 1997).
Spectral analysis showed the source to be relatively soft, with a
power law index of around 2.7 (although a power-law is not the best fit
model).  A Fourier analysis did not reveal any significant
periodicities, with the authors lamenting an increase in spectral
noise at frequencies below 0.1 Hz.
 
The companion star first suggested by Clark, Remillard \& Woo (1996)
was observed by Charles, Southwell \& O'Donoghue (1996)  optically
1996 January.  These authors determined that the
B1-2 star of magnitude 14.2 proposed as the companion showed a strong
IR excess and Balmer lines and a reddening typical of an OB star in
the SMC, thus strengthening the association of the X-ray source with
the SMC.  They also argue that the luminosity and companion type
indicates that the X-ray source is a neutron star (Coe et al. 1997).
However, Clark et al. (1997) hypothesize that the system could harbor
a black hole based on the e-folding X-ray decay time of 44 days, the
rather soft spectrum, and the lack of any neutron star rotation period
in the X-ray analysis.  

We have performed a reanalysis of the
ROSAT/PSPC data and coupled it with hard X-ray observations by the
CGRO/BATSE instrument.  In Section 2, we present evidence for X-ray
emission pulsed at a 22.067 second period which definitively establishes
the X-ray source as a neutron star.  We show the frequency
history for RX J0117.6-7330 during this 100 day outburst which reveals
an extremely large average frequency derivative of $8.9\times10^{-11}$
Hz s$^{-1}$ corresponding to a spin-up time scale of 16 years. 
The frequency derivative peaked at $1.2\times10^{-10}$ Hz
s$^{-1}$, with the pulse frequency increasing by 1.8\% during the BATSE
observations. The broad-band X-ray pulse shapes and pulsed flux are
calculated in Section 3.  Section 4 summarizes our findings and
discusses RX J0117.6-7330 in the context of the Be class of HMXB's.

\section{Periodicity Search}

RX J0117.6-7330 was 5 arcminutes from the center of the PSPC
field-of-view during an 8985 second exposure taken from MJD 48895.7
- 48897.6 (MJD = JD - 2400000.5). 
We determine a total source count rate of $4.43\pm0.03$
cts/sec from 0.1-2.4 keV.  
Photons in a circle
of radius 1 arcminute surrounding the source position J2000 RA,Dec:
(01 17 36, -73 30 00) were extracted and barycentered using standard
FTOOLS software.  A total of 33989 photons were available for timing
analysis.
 
These photons were collected into 5 msec bins over the full
length of the observation, a time span of 162 ksec.  An FFT of the
resultant time series was then calculated, sensitive to periods in the
range from 10 msec to 81000 seconds with the power per channel
normalized to unity using the average power for all frequencies above 0.01 Hz.
No frequency derivatives were
included at this point of the analysis.  A peak of power 30, normalized
as above, at a
frequency of 0.090825(2) Hz was evident in this analysis which 
warranted further study despite increased noise due to the 
complicated ROSAT
exposure induced window function and spacecraft wobble.

Verification of the pulsed signal comes from an archival search of
data from the BATSE instrument on the Compton Gamma-ray
Observatory.  BATSE is capable of nearly continuous monitoring of hard
X-ray sources using both the earth occultation technique (Harmon et
al. 1992) and timing techniques (Bildsten et al. 1997).  
A search of archival BATSE FFT results identified a
possibly related outburst some 60 days after the 1992 ROSAT
observation at a frequency of 0.0914825(2) Hz, with a frequency
derivative of $9.7(2) \times 10^{-11}$ Hz s$^{-1}$ consistent with the direction
of the SMC.  With follow-up Epoch-folding based searches of the BATSE data
during the ROSAT observation, it became apparent that the originally
detected frequency was the second harmonic of the pulse frequency.
From six days of BATSE data centered on the ROSAT observation, a
barycentric pulse frequency of 0.045316682(55) Hz (MJD 48896.65) and
frequency derivative of $9.81(8)\times10^{-11}$ Hz s$^{-1}$ were determined.
Folding the ROSAT data with this frequency derivative gives a peak power
at 0.0453168(3) Hz, consistent with the BATSE pulse period.  
Figure 1 shows the resultant ROSAT power spectrum calculated using the
Z$^{2}_{2}$ statistic (Buccheri et al. 1983) over a narrow frequency 
range utilizing the
above-stated frequency derivative.  Also plotted is the same statistic
for the BATSE data at frequencies near the ROSAT signal.  

The results of searching the BATSE DISCLA channel 1 data (20-50 keV,
1.024s resolution) from 1992 August 16 (MJD 48850) to 1993 January 12
for pulsations from RX J0117-7730 are presented in Fig. 2. These
searches were performed in six day intervals, using an epoch-folding
based search (see Bildsten et al. 1997) which used only the
first and second harmonic of the pulse profile, and incorporated
a search in both pulse frequency and frequency derivative. For
intervals where pulsations were detected the pulse frequency and
frequency derivative are shown. The frequency derivative peaks at
$1.2\times 10^{-10}$ {\rm Hz~s}$^{-1}$ 25 days before the 1992 ROSAT
observations. The pulse frequency increases by 1.8\% during the
outburst.  The signal is present for approximately 100 days, starting
about 34 days before the ROSAT/PSPC observation.

\section{Broad-band Pulse Profile and Flux}

Using the measured frequency and frequency derivative, we 
can construct the pulse profile for RX J0117.6-7330 for both
soft and hard X-ray energies.  Figure 3 shows both the ROSAT/PSPC
pulse profile for 1992 Sep 30 - Oct 2 (MJD 48895-48897)and the
CGRO/BATSE profile for MJD 48893.65 - 48899.65.  Both datasets use
the pulse phase model based on the BATSE data.
The optimal ROSAT frequency and frequency derivatives are slightly
different.  While the very high frequency derivative, coupled with the
long integration times (6 days for BATSE, 2 days for ROSAT) make
absolute timing comparisons slightly problematic, there is no evidence
for a loss of coherence in the BATSE folding and the pulse phases from
the two instruments should be directly comparable.   Figure 3 shows the
pulse shapes for both energy ranges using an epoch at phase zero of
MJD 48896.65.  An extra phase offset of 0.14 was added in order to make
the BATSE minimum correspond to phase zero.  
The overall profile shapes are similar.
The peaks and minima of the lightcurves are generally in phase, but the primary
peak in the ROSAT energy range becomes the secondary peak in the BATSE
range.  Similarly, primary and secondary minima are interchanged.

We have
tested our timing analysis methods using contemporaneous ROSAT/PSPC and
CGRO/BATSE observations of PSR 1509-58.  In this case, we find pulse
profiles with shapes and radio-phase offsets consistent with previously
published results(Greiveldinger et al. 1995, Ulmer et al. 1993). 
The overall pulse shape for RX J0117.6-7330 in both energy ranges
 are similar, so it is not out of the quesion that
the shape as a whole has simply shifted. 
Apparent phase shifts of simple profiles in
different energy bands have been previously observed.  For example,
the 1.2-2.3 keV and 18.4-27.5 keV profiles of GS 0834-430
observed with Ginga by Akoi et al. (1992) show a complex evolution with
energy.  It would be somewhat coincidental, however, for the phase shift
to be such that the minima and maxima still coincide.  At this point, we
consider the peaks to be aligned, with the relative strengths to be changing.

From these pulse profiles one may determine the pulsed flux and
pulsed fraction.  
Using XSPEC, we calculate a total flux of $5.1\pm0.3 \times 10^{-11}$ ergs
cm$^{-2}$ s$^{-1}$.  Similar to Clark et al. 1997, we find the best fit
to the ROSAT spectrum is a combination of power-law and bremsstrahlung
or blackbody although a straight power-law fit of index $2.65\pm0.07$
is not much worse.  Of the 33989 total counts extracted for the light
curve, 3829 comprise the pulsed excess giving a pulsed percentage of
$11.3\pm2.3$\%.  This corresponds to a total pulsed flux in the 0.2 to
2.5 keV band of $5.6\pm1.7\times10^{-12}$ ergs cm$^{2}$ s$^{-1}$.
In the BATSE energy range, the phase-averaged pulsed fraction is more
difficult to assess.  An occultation analysis of RX J0117.6-7330
detects a clear signal over the same time frame as the epoch-folding
analysis.  However, source confusion could significantly contribute to
the detected flux.  For a 20 day period around the time of the ROSAT
observation, the average total flux is $0.012\pm0.02$ cm$^{-2}$ s$^{-1}$.
This corresponds to an energy flux of $2.3\pm0.4 \times 10^{-10}$ ergs
cm$^{-2}$ s$^{-1}$ (assuming a power law index 3.0, 20-100 keV).  
The BATSE pulsed
spectrum is best fit by a thermal bremsstrahlung model with
temperature $18 \pm 3$ keV.  The integrated 20-70
keV pulsed flux is $1.8\pm0.2 \times 10^{-10}$ erg cm$^{-2}$ s$^{-1}$ ( $7.8
\times10^{37}$ (D/60 kpc)$^{2}$ ergs s$^{-1}$ ).  These values provide us
with a lower limit to the pulsed fraction in the 20-70 keV range of
78\% (a lower limit since the measured occultation flux is considered
to be an upper limit to total emission due to possible source confusion
and the occultation analysis went up to 100 keV).  The
directly measured flux in the 0.2-2.5 and 20-70 keV bands alone
is then at least $2.3\pm0.2 \times 10^{-10}$ erg cm$^{2}$ s$^{-1}$
($1.0\pm0.1 \times10^{38}$ (D/60 kpc)$^{2}$ ergs s$^{-1}$). 

\section{Discussion}

X-ray pulsations at a period of 22.07 seconds from the bright X-ray
transient RX J0117.6-7330 have been detected in both the 0.1-2.4
and 20-70 keV energy bands.  This confirms the identity of the X-ray
source as a neutron star rather than a black hole.  Transient X-ray
pulsars are typically found in Be systems which is consistent with,
and supports the identification of, the proposed optical counterpart.

The CGRO/BATSE detection allows long-term monitoring of the outburst
from this source.  The hard X-rays are detectable for over 100 days
starting 34 days before the ROSAT observation began.  A large average
spin-up is present over the duration of the outburst resulting in a
1.8\% change in frequency.  The peak frequency changes appear 10-15
days before the ROSAT observation and approximately 20 days after the
outburst began.  Some of the frequency derivative may be caused
by the binary orbit.  Some of the variations seen in Figure 2 near the
peak of the frequency derivative may be an orbital signature.  Such
variations are weak, however, compared to the overall accretion induced
changes in frequency.  

The high intrinsic spin-up rates imply that an accretion
disk is present about the neutron star, as is generally seen in the
`giant' or type I outbursts of Be/X-ray pulsars (Bildsten et al. 1997).  
The measured
luminosity is at least $1.0 \times 10^{38}$ erg s$^{-1}$ during the
ROSAT observation in the combined 0.2-2.5 and 20-70 keV bands.  
The peak frequency derivative was about 15\%
higher than during the ROSAT observation.  However, we have no
data in the energy range from 2.5 - 20 keV.  With a complicated ROSAT
X-ray spectrum and a changing pulse fraction, is difficult to
extrapolate our results to this energy range.  However, it is likely
that the luminosity in this range is comparable to that measured in
the 0.2-2.5 and 20-70 keV ranges.  Thus the peak luminosity, when
adjusted for higher frequency derivative and 2-20 keV emission is $\geq
1 \times 10^{38}$ erg s$^{-1}$, and was
probably higher than the conventional Eddington limit.  
This is consistent with 
the trend for Magellanic cloud binaries to be much brighter on average
than their galactic counterparts probably due to the absence of metals
which supply accretion inhibiting absorption
(Clark et al. 1978; van Paradijs \& McClintock 1995).  
From the peak spin-up rate and standard
accretion theory (Bildsten et al. 1997 and references therein),
we can obtain a lower limit on the peak luminosity
of around $2.5\times10^{38}$ erg s$^{-1}$. This is consistent with
the mean luminosities of X-ray binaries in the SMC (van Paradijs \&
McClintock 1995). It is inconsistent with
the source being a galactic object.

The pulse profiles in the two energy bands are similar in that they both
have a double peaked structure.  However, the 
main and secondary peaks are interchanged.  
X-ray binaries typically have pulse profiles which are often strongly
energy dependent (White, Swank \& Holt 1983).  In this case, the
double peaked lightcurve in both energy bands show the same
peak-to-peak separation of 0.5.
The pulsed fraction increases from 11\% in soft X-rays to at
least 78\% in hard X-rays.  If the overall morphology of the pulses is
indeed the same, with the exception of the relative strengths of the two peaks,
this may indicate a high magnetic field since at
energies above the cyclotron energy, the pulse shape is expected to
change significantly (e.g. see Sturner and Dermer 1994).

{\bf Acknowledgements } 

This project made use of software and data provided by the High-Energy
Astrophysics Archival Research Center (HEASARC) located at Goddard
Space Flight Center.  This work was supported at Caltech in part by NASA
NAG 5-3239.

\newpage

\section{References}
\vskip 5pt
Aoki, T., et al. 1992, PASJ, 44, 641
\vskip 5pt
Bildsten, L., et al. 1997, ApJS, 113, 367 
\vskip 5pt
Buccheri, R., et al. 1983, A\&A, 128, 245
\vskip 5pt
Charles, P.A., Southwell, K.A., \& O'Donoghue, D. 1996, IAUC 6305
\vskip 5pt
Clark, G., Doxsey, R., Lie, F., Jernigan, J.G. \& van Paradijs, J. 1978, ApJ, 221, L37
\vskip 5pt
Clark, G., Remillard, R., \& Woo, J. 1996, IAUC 6282
\vskip 5pt
Clark, G., Remillard, R., \& Woo, J. 1997, ApJ, 474, L111 
\vskip 5pt
Coe, M.J., Buckley, D.A.H., Charles, P.A., Southwell, K.A., \& Stevens, J. B 
1998, MNRAS, 293, 43C
\vskip 5pt
Harmon, A. et al. 1993, in ``Compton Gamma-Ray Observatory'', AIP
Conf. Proceedings 280, (AIP: New York), 314
\vskip 5pt
Greiveldinger, C., Caucino, S., Massaglia, S., Ogelman, H. \& Trussoni,
E. 1995, ApJ, 454, 855
\vskip 5pt
Sturner, S.J. \& Dermer, C.D. 1994, A\&A, 284, 161
\vskip 5pt
Ulmer, M.P., et al. 1993, ApJ, 417, 738
\vskip 5pt
van Paradijs, J. \& McClintock, J.E. 1995, in "X-ray binaries", ed.
Lewin, W.H.G., van Paradijs, \& van den Heuvel, E.P.J., Cambridge
University Press, p. 113
\vskip 5pt
White, N.E., Swank, J.H., \& Holt, S.S. 1983, ApJ, 270, 711
\vskip 5pt 
White, N.E., Giommi, P. \& Angelini, L. 1995, BAAS, 185
\newpage

\centerline{\psfig{figure=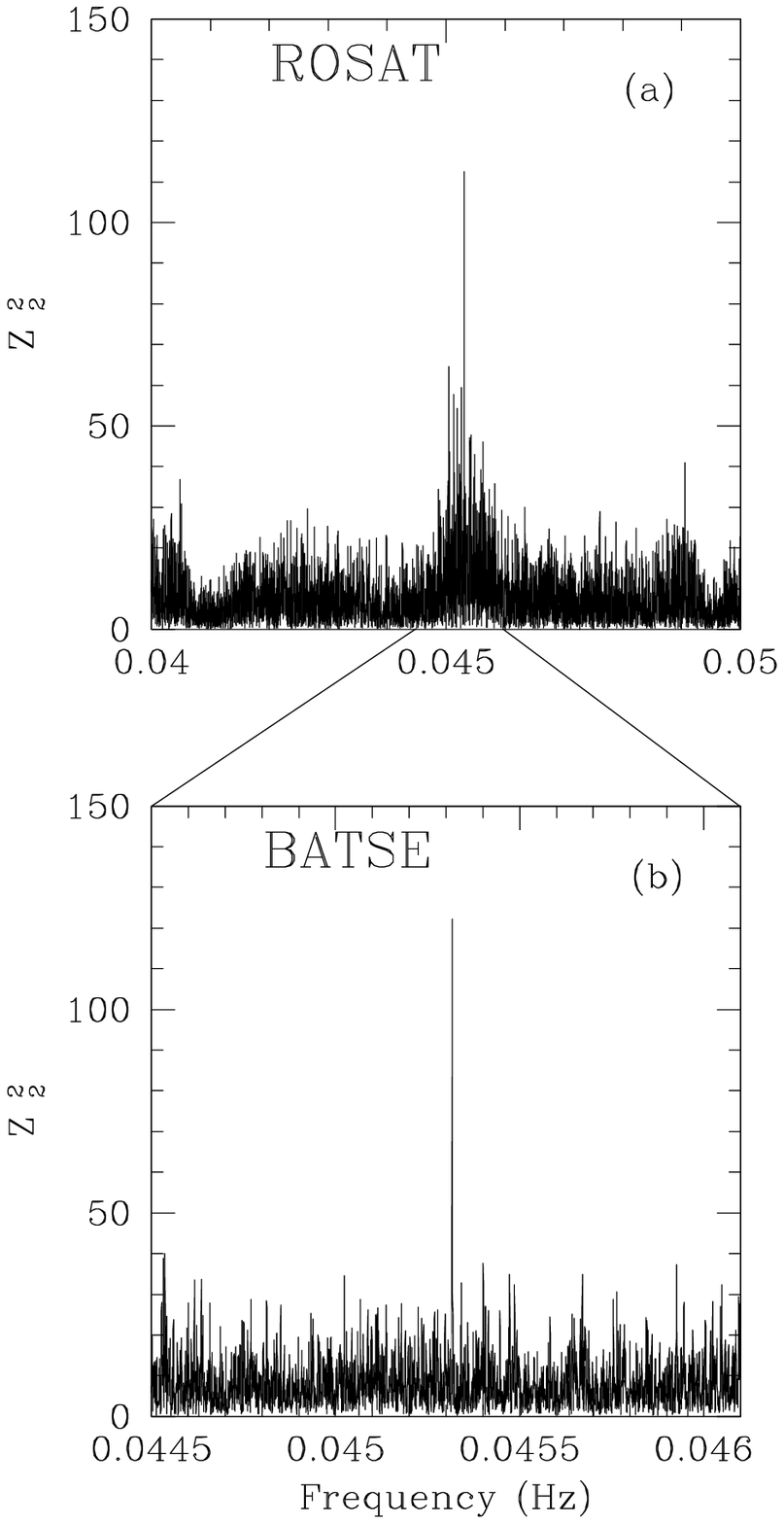}}
{\bf Figure 1}: The ROSAT power distribution encompassing the first
harmonic of the pulsed frequency.  The BATSE distribution is for a narrow 
range around the ROSAT detection frequency.   
\vskip 12pt
\newpage
\centerline{\psfig{figure=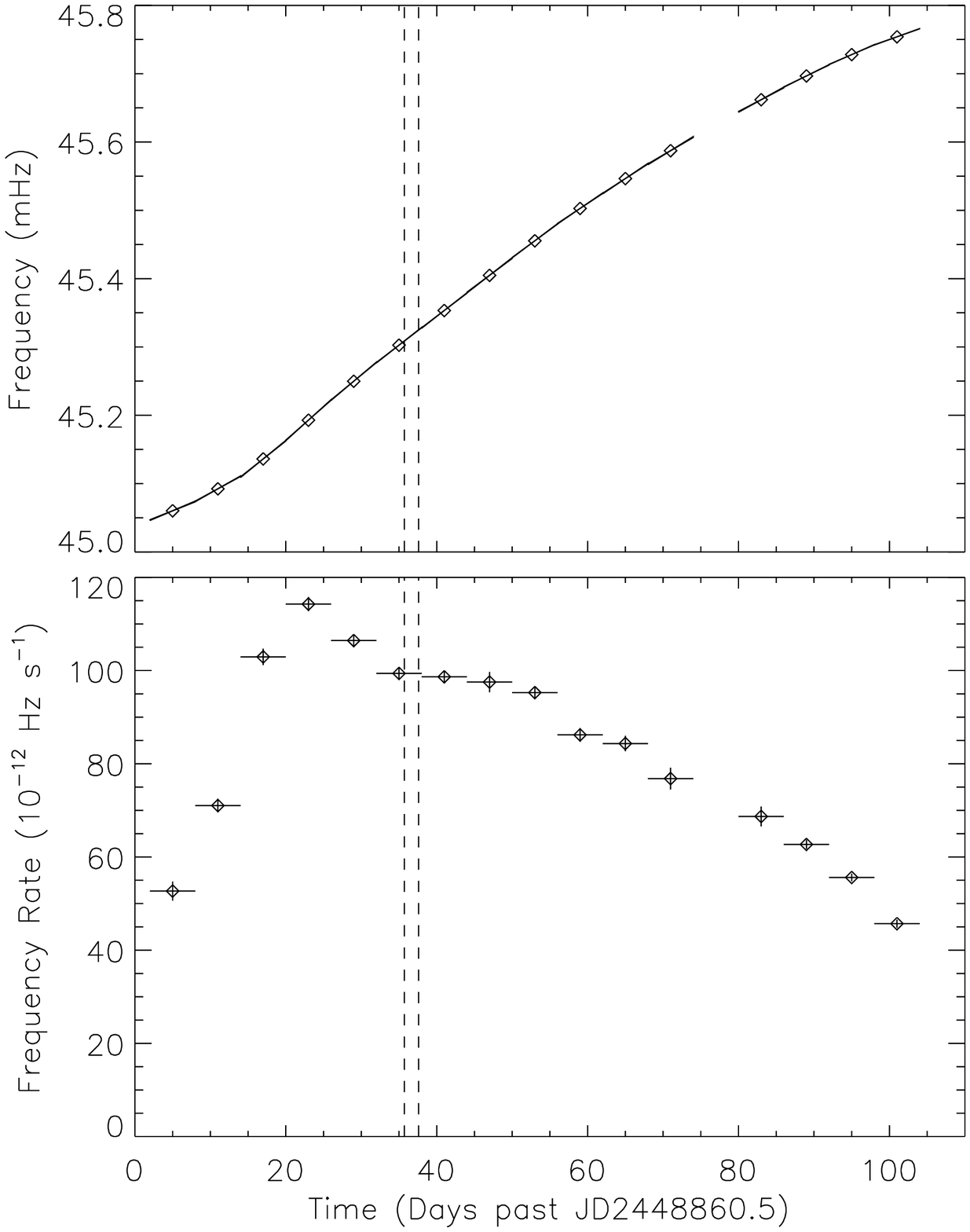}}
{\bf Figure 2}: BATSE pulse timing analysis of RX J0117.6-7330.  The
top panel is the frequency history and the bottom panel the frequency
rate history.  Both plots use the best fit values for 6 day intervals
with the ROSAT observation date shown by the dashed line.
\vskip 12pt

\newpage
centerline{\psfig{figure=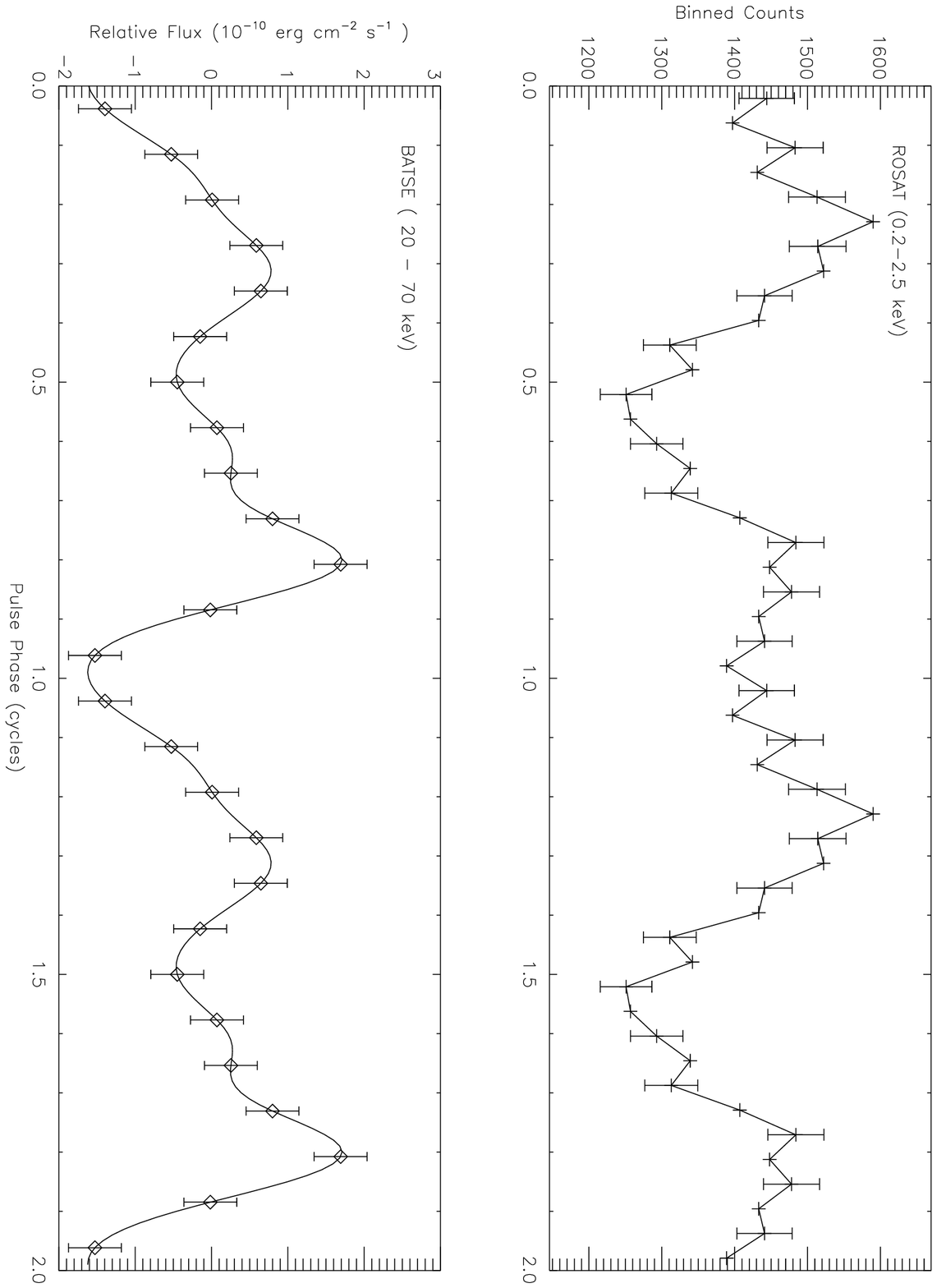}}
{\bf Figure 3}: The folded light curves for ROSAT 0.1-2.4 keV (top)
and BATSE 20 - 70 keV (bottom) data. The BATSE profile is limited to six
Fourier harmonics, resulting in the smooth shape.  Flux errors
are given for a set of approximately independent phases.

\end{document}